\documentstyle[11pt]{article}
\def\beq{\begin{equation}}
\def\eeq{\end{equation}}

\begin{document}
\ .
\vskip 3.2cm
\noindent{\Large{\bf Blocking of inter-subspace tunneling by 
intra subspace inelastic scattering. }}

\vspace{1.cm}
\hangindent=40pt
T.P. Pareek$^*$, A.M. Jayannavar$^*$ and N. Kumar$^{**}$

\vspace{.2cm}
\hangindent=40pt
$^*$Institute of Physics, Sachivalaya Marg, Bhubaneswar 751005, India

\hangindent=40pt
$^{**}$ Raman Research Institute, Bangalore 560080, India

\vskip 0.7cm

\hangindent=20pt {\bf Abstract :}
In recent years the notion of intrinsic decoherence and dephasing of a
particle interacting with its environment is being investigated
intensively. This has an important bearing on a plausible causal
connection between incoherent c-axis resistivity and high-temperature
superconductivity. In our work we study the tunnel
supression and incoherent motion of a particle tunneling between two
sites. The bosonic excitations of the environment 
are coupled to only one site inducing on-site spin flips. 
We show that this on-site spin flip scattering makes the tunnel motion
incoherent. In the high-temperature limit incoherent rate or hopping
rate has been calculated. We also briefly discuss the renormalization
of the effective tunneling by environmental coupling at zero
temperature following Wegner's renormalization group procedure.
\vskip .7cm

\hangindent=20pt
{\bf Keywords:} tunneling, incoherence, spin-flip, renormalization 

\vskip 0.7cm

\hangindent=1.2cm
{\bf PACS Nos:} 05.40.+j, 05.60.+w , 63.35.+a , 64.60.Ak.

\vskip 1.cm

One of the most interesting and physically realizable low-dimensional
system is a strongly correlated metal, with the possibility of
electrons
escaping into the third direction by a weak tunneling. Generalization
to a d-dimensional system with escape into (d+1)th dimension is
obvious, when d=0 (quantum dot), d=1(quantum wire) or d=2 (quantum
sheet). For an important realization one may consider a class of
high-$T_c$ layered cuprates, where metallic 2-dimensional
CuO$_2$-sheets are weakly coupled across the spacer layers of various
oxides of Ca, Sr, Bi etc \cite{gins,clark}. 
It has been proposed earlier that the normal 
state c-axis resistivity ($\rho_c$) in the highly anisotropic systems is
incoherent and is controlled by the ab-plane resistivity ($\rho_{ab}$) 
\cite{amj,amj1,leg}. 
Indeed in the best single crystal samples both $\rho_c$ and $\rho_{ab}$
are linear in temperature, from $T_c$ ($10^2$K) right up to $10^3$K 
(far above the Debye temperature) with $\rho_c$ $\propto$ $\rho_{ab}$.
This has been attributed to the blocking of already weak inter-planer
tunneling ($t_\perp$) by a strong intra-planer scattering rate 
($1/\tau_{\parallel}$ of electronic origin). This ``quantum Zeno
efect''
was discussed interms of well known spin-boson model Hamiltonian
where the tunneling rate is cutoff by its coupling to a bosonic
bath after Caldeira and Leggett \cite{clark,rmp}. In this work we have reexamined this
blocking effect through a simple spin-boson model which is {\bf
qualitatively different} from the two-state Caldeira-Leggett model,
and is more directly related to the problem of blocking of inter-planer
tunneling by blocking of intra-planer scattering. Thus, whereas in the
usual spin-boson model, the blocking is due to the adiabatic overlap
of the two bosonic ground states (displaced Harmonic oscillators) for
the particle in two states, that multiplies the electronic tunneling
matrix element, in the present case it is the decoherence due to the
intra-planar incoherent dynamics that cut-off the inter-planar
coherent tunneling - a truely Quantum Zeno effect.

We have considered the model Hamiltonian
\begin{equation}
  \label{eq:1}
  H=H_{S}+H_{B}+H_{SB} ,
\end{equation}
\noindent where
\begin{eqnarray}
\label{eq:2}
H_{S}=V\sum_{\sigma}(c^{\dagger}_{1\sigma}c_{2\sigma}+
c^{\dagger}_{2\sigma}c_{1\sigma}), \\
H_{B}=\sum_{q}\hbar\omega_{q}(a_{q}^{\dagger}a_{q}+1/2),\\
H_{SB}=\sum_{q}\alpha_{q}(a_{q}+a_{q}^{\dagger})
(c^{\dagger}_{1\uparrow}c_{1\downarrow}+
c^{\dagger}_{1\downarrow}c_{1\uparrow}).
\end{eqnarray}

Here $H_S$ describes the tunneling between the two sites 1 and
2 and $c_{i\sigma}$, $c_{i\sigma}^{\dagger}$ are the annihilation and
creation operators of the tunneling particle with $i$
and $\sigma$ being site and spin index.
$H_B$ is the harmonic bath Hamiltonian ,$a_q$,$a_{q}^{\dagger}$ are the 
bosonic annihilation and creation operators, and $H_{SB}$ couples the bath to
site 1 inducing the onsite spin flips. Thus, $H_S$ simulates the
inter-planar tunneling (conserving spin $\sigma$), and $H_{SB}$
models the intra-planar dynamics rendered incoherent by coupling to
the bath $H_B$. The question now addresed is how the on-site
incoherent spin-flip dynamics blocks the tunneling rate
1$\rightarrow$2 between the sites 1 and 2. For this
we first derive the quantum Langevin equations for the system
variables and obtain effective incoherent tunneling 
rate in the high 
temperature limit. Finally we describe a renormalization group
procedure for calculating  effective tunnel matrix element at zero temperature.

{\bf Quantum Langevin Equations; dynamics in the high temperature limit }

To obtain the quantum Langevin equations for the system
variables $c_{i\sigma}$ and $c_{i\sigma}^{\dagger}$. We use the
anticommutation relations \{$c^{\dagger}_{i\sigma},c_{j{\sigma^\prime}}$\} =
$\delta_{ij}\delta{\sigma{\sigma^\prime}}$, commutation relations
$[a_{k},a_{k^\prime}^{\dagger}]=\delta_{k{k^\prime}}$ for the bath
variables and the Heisenberg equation of motion for any operator
A, namely ${dA/dt} = \frac{1}{i\hbar}{\left[A, H\right]}$. We get

\begin{eqnarray}
\label{eq:3}
i\hbar\frac{d(c^{\dagger}_{i\alpha}c_{j{\beta}})}{dt}&=&
V\sum_{\sigma}c^{\dagger}_{i\alpha}c_{2\sigma}\delta_{j1}\delta_{\sigma\beta}
+V\sum_{\sigma}c^{\dagger}_{i\alpha}c_{1\sigma}\delta_{j2}\delta_{\sigma\beta} 
-V\sum_{\sigma}c^{\dagger}_{1\sigma}c_{j\beta}
\delta_{i2}\delta_{\sigma\alpha} \nonumber \\
&-&V\sum_{\sigma}c^{\dagger}_{2\sigma}c_{j\beta}
\delta_{i1}\delta_{\sigma\alpha} 
+\sum_{q}\alpha_{q}(a_{q}+a_{q}^{\dagger})
c_{i\alpha}^{\dagger}c_{1\downarrow}
\delta_{j1}\delta_{\beta\downarrow} \nonumber \\
&-&\sum_{q}\alpha_{q}(a_{q}+a_{q}^{\dagger})c_{1\uparrow}^{\dagger}c_{j\beta}
\delta_{i1}\delta_{\alpha\downarrow} \nonumber \\
&+&\sum_{q}\alpha_{q}(a_{q}+a_{q}^{\dagger})c_{i\alpha}^{\dagger}c_{1\uparrow}
\delta_{j1}\delta_{\beta\downarrow}
-\sum_{q}\alpha_{q}(a_{q}+a_{q}^{\dagger})c_{1\downarrow}^{\dagger}c_{j\beta}
\delta_{i1}\delta_{\alpha\uparrow}.
\end{eqnarray}
Here $i$ , $j$ and $\alpha$, $\beta$ takes on values 1,2 and
$\uparrow$,
$\downarrow$ respectively with all possible combinations.
\begin{equation}
\label{eq:4}
\frac{da_q}{dt}= -\frac{i}{\hbar}\alpha_{q}
(c^{\dagger}_{1\uparrow}c_{1\downarrow}
+c^{\dagger}_{1\downarrow}c_{1\uparrow})- i \omega_{q}a_{q},
\end{equation}
\begin{equation}
\label{eq:5}
\frac{da_{q}^{\dagger}}{dt}= \frac{i}{\hbar}\alpha_{q}
(c^{\dagger}_{1\uparrow}c_{1\downarrow}
+c^{\dagger}_{1\downarrow}c_{1\uparrow})+ i \omega_{q}a_{q}^{\dagger}.
\end{equation}
Eqs. (\ref{eq:4}) and (\ref{eq:5}), being linear, can be readily
solved. We then substitute the formal solutions of
$a_{q}^{\dagger}(t)$ and $a_{q}(t)$ (which involve initial values of 
variables $a_{q}^{\dagger}(0)$ and $a_{q}(0)$ at time $t$=0) in
eqn(\ref{eq:3}). If one assumes the Ohmic spectral density for bath variables, 
i.e., $\rho(\omega)=\frac{\pi}{2}\sum_{q}\frac{4\alpha_{q}^{2}}{\hbar^2}
\delta(\omega-\omega_q)=\alpha\omega$, $\alpha$ being the coupling
constant or Kondo parameter, we get Markovian quantum Langevin
equations (for the details see \cite{pareek}).
\newpage
\begin{eqnarray}
\label{eq:8}
i\hbar\frac{d(c^{\dagger}_{i\alpha}c_{j{\beta}})}{dt}&=&
V\sum_{\sigma}c^{\dagger}_{i\alpha}c_{2\sigma}\delta_{j1}\delta_{\sigma\beta}
+V\sum_{\sigma}c^{\dagger}_{i\alpha}c_{1\sigma}\delta_{j2}\delta_{\sigma\beta} 
-V\sum_{\sigma}c^{\dagger}_{1\sigma}c_{j\beta}
\delta_{i2}\delta_{\sigma\alpha} \nonumber \\
&-&V\sum_{\sigma}c^{\dagger}_{2\sigma}c_{j\beta}
\delta_{i1}\delta_{\sigma\alpha} +
\left\{F(t)-\frac{i\eta}{\hbar}\left[c^{\dagger}_{1\downarrow}c_{2\uparrow}
-c^{\dagger}_{2\uparrow}c_{1\downarrow}+c^{\dagger}_{1\uparrow}c_{2\downarrow}
- c^{\dagger}_{2\downarrow}c_{1\uparrow}\right] \right\}  \nonumber \\
&&\left[c_{i\alpha}^{\dagger}c_{1\downarrow}
\delta_{j1}\delta_{\beta\downarrow} 
-c_{1\uparrow}^{\dagger}c_{j\beta} 
\delta_{i1}\delta_{\alpha\downarrow} 
 +  {c_{i\alpha}^{\dagger}c_{1\uparrow}}
\delta_{j1}\delta_{\beta\downarrow}
-c_{1\downarrow}^{\dagger}c_{j\beta}
\delta_{i1}\delta_{\alpha\uparrow}\right],
\end{eqnarray}
where $F(t)$ is 
\begin{equation}
F(t)=\sum_{q}\alpha_{q}(a_{q}(0)e^{-i\omega_{q}t}+
a_{q}^{\dagger}(0)e^{i\omega_{q}t}).
\end{equation}
As the  operators $a_{q}(0),a_{q}^{\dagger}(0)$ of the bath
are distributed in accordance with the statistical equilibrium distribution 
for given temperature $T$, $F(t)$ is referred as Langevin operator
noise term.
The statistical
properties of $F(t)$ can be obtained using the equlibrium
distribution for bath varaibles togther with the
Ohmic spectral density. Owing to the operator nature of 
the random Langevin force $F(t)$,
it is difficult to solve for the expectation values of site occupancy
using equations (\ref{eq:8}).
However in the high temperature limit (made precise in the ref. \cite{pareek}),
one can treat $F(t)$ as a classical c-number random variable.
One can readily verify that in the
classical limit 
taking $\hbar\rightarrow 0$, the nonequal
time commutator of $F(t)$ vanishes and the autocorrelation of
the Gaussian random force $F(t)$ becomes
\begin{equation}
\label{scfc}
\langle{{F(t)}{F(t^{\prime})}}\rangle=\eta kT\delta(t-t^{\prime}),
\end{equation}
\noindent where $\eta$ is the dissipation coefficient and is related to
Kondo parameter $\alpha$
($\eta=(\hbar \alpha/2$) \cite{pareek}. Henceforth we set $\hbar$
to be unity.
With the use of Novikov's theorem \cite{pareek} for the functionals of Gaussian
variables we can compute the expression for the averaged quantum
expectation value for the occupation probability of a particle on site
2, $n_{2}(t)$=$<\sum_{\sigma}c^{\dagger}_{2\sigma}c_{2\sigma}>$
($<$...$>$ brackets denotes the average over the stochastic variable
$F(t)$), subject to the initial condition that the system was prepared 
initially at t=0 on the site 1. For this we have solved the set of
coupled linear equations and the final result is
\begin{eqnarray}
\hspace{-5cm} n_{2}(t) &\equiv& <\sum_{\sigma}c^{\dagger}_{2\sigma}c_{2\sigma}> 
\nonumber \\
&=&\frac{4V^{2}}{b}
\left\{ \frac{e^{-t \frac{a+b}{2}}}{a+b} 
- \frac{e^{-t \frac{a-b}{2}}}{a-b}\right\}
 +\frac{1}{2}
\end{eqnarray}
\noindent where $a$=$\eta k_{B}T$ and
$b$=$\sqrt{(\eta k_{B}T)^{2}-16 V^{2}}.$

In the absence of environment coupling ($\eta$=0), the particle executes
the coherent tunneling oscillations between the two sites with
frequency 2$V$, namely $n_{2}(t)$=$(1/2)(1-cos(2Vt))$.
As the coupling (or temperature) is increased thermally induced onsite
spin-flip scattering destroys the coherence and $n_{2}(t)$ approaches
the equlibrium value 1/2 in the asymptotic time limit.
 The rate of tunneling decreases rapidly as
$\eta$ (the strength of coupling to bath) increases. Indeed, the
exponetial in the expression for $n_{2}(t)$ can be approximated for 
$\eta k_{B}T >> V$
to give the incoherent tunneling rate $\sim$ (4$V^2$/$\eta k_{B}T$), that 
decreases monotonically with increasing $\eta$.
In this regime $n_{2}(t)$ exhibits no oscillations and approaches
monotonically the value 1/2. The particle hops randomly (incoherently) with
no fixed tunneling period and the motion becomes overdamped. 
This is in agreement
with the result obtained by Kumar and Jayannavar \cite{amj}. Thus we have
established that onsite spin flip scattering decreases the incoherent tunnel rate.

{\bf Flow equations for effective tunnel matrix element.}

In this section using continuous unitary transformation introduced
recently by Wegner \cite{wegner}, we obatin flow equations for the coupling
parameters in the Hamiltonian H . In this approach the Hamiltonain is
diagonalised by continuous  infinitesimal unitary transformation
starting from the original Hamiltonain, H(l=0)= H and terminating
with a diagonal Hamiltonain with renormalized coupling constants as 
$l\rightarrow\infty$. Here $l$ is the flow parameter labelling the
tunneling rate $V(l)$ and coupling constant $\alpha_{k}(l)$ under the
transformation. The flow equations can be written in a differential form
\begin{equation}
\frac{dH}{dl}=\left[\eta(l),H(l)\right], \,\,\,\, H(l=0)=H ,
\end{equation}
\noindent where $\eta$ is the generator of the infinitesimal unitary
transformation, it is an anti-hermitain operator that depends on H and 
therefore implicitly on the flow parameter $l$. 
Wegner proposed to choose $\eta(l)=[H_{d}(l),H(l)]$, where
$H_{d}(l)$ is the appropriate diagonal part of $H(l)$. However, there
are several possibilites to choose $\eta$ so that $H(\infty)$ becomes 
diagonal. We have made the following ansatz for  $\eta$,
\begin{eqnarray}
\eta(l)= -\sum_{kx}\eta_{kx}(l)(a_{k}+a_{k}^{\dagger})(c^{\dagger}_{2\uparrow}
c_{1\downarrow}+ c^{\dagger}_{2\downarrow}
c_{1\uparrow}-c^{\dagger}_{1\uparrow}
c_{2\downarrow}- c^{\dagger}_{1\downarrow}
c_{1\uparrow})  \nonumber \\
+\sum_{ky}\eta_{ky}(l)(a_{k}^{\dagger}-a_{k})(c^{\dagger}_{1\uparrow}
c_{1\downarrow}+ c^{\dagger}_{1\downarrow}c_{1\uparrow}) 
+\sum_{kz}\eta_{kz}(l)(a_{k}^{\dagger}-a_{k})(c^{\dagger}_{2\uparrow}
c_{2\downarrow}+ c^{\dagger}_{2\downarrow}c_{2\uparrow}),
\end{eqnarray}
where $\eta_{kx}$,$\eta_{ky}$ and $\eta_{kz}$ are coefficients to
be determined. The flow equations for the parameter of the original
Hamiltonian generates interactions not contained in the original
Hamiltonian which are quadratic in the bath operators.
We have neglected them, as one usually expects
them
to be unimportant for low lying excitation of the systems at T=0 \cite{mielke}.
Following closely the procedure given in \cite{mielke} we finally obtain the
flow equations of the effective tunnel matrix elemnt $V(l)$, for Ohmic
spectral density of bath (the details will be published elsewhere) and
is given by

\begin{eqnarray}
\frac{dV(l)}{dl}&=&-\alpha \int_{0}^{\omega_c} \omega^{2} exp\left(2\int_{0}^{l}
\frac{4V^{2}(l)-(\omega)^{2}}{V(l)} \omega dl \right) d\omega.
\end{eqnarray}
\noindent where $\omega_c$ is an upper cutoff freequency of harmonic
bath.
We have verified that $V(0) > 0$ initially, $V(l)$ decreases
monotonically to zero as $l\rightarrow\infty$,
indicating the complete
suppression of tunneling between two sites.
This is a simple case of orthogonal catastrophe. Thus at zero
temperature on-site spin flip scattering suppresses the tunneling.

In conclusion we have shown that onsite spin-flip scattering induced
by environment makes tunnel motion incoherent. The intra-site
spin-flip scattering blocks the inter-site tunneling. The calculated
incoherent tunnel rate in the high temperature limit is in agreement
with earlier known results. 
At zero temperature the tunnel motion is suppressed.

\end{document}